\documentclass[prb,twocolumn,showpacs,showkeys,groupedaddress,preprintnumbers,amsmath,amssymb]{revtex4}
\usepackage {amssymb}
\usepackage {amsmath}
\usepackage{graphics,amssymb}
\usepackage{natbib}
\usepackage{revsymb}

\usepackage{graphicx}
\usepackage{dcolumn}
\usepackage{bm}

\begin{document}


\title{Modeling of the transient interstitial diffusion of implanted atoms during low-temperature annealing of silicon substrates\\}

\author{O. I. Velichko} \email[]{velichkomail@gmail.com}
\affiliation{Department of Physics, Belarusian State University of
Informatics and Radioelectronics, 6 P. Brovki Str., Minsk 220013,
Belarus
\\}

\author{A.P. Kavaliova} 
\affiliation{Department of Physics, Belarusian State University of
Informatics and Radioelectronics, 6 P. Brovki Str., Minsk 220013,
Belarus
\\}


\begin{abstract}
It has been shown that many of the phenomena related to the
formation of ``tails'' in the low-concentration region of
ion-implanted impurity distribution are due to the anomalous
diffusion of nonequilibrium impurity interstitials. These
phenomena include boron implantation in preamorphized silicon, a
``hot'' implantation of indium ions, annealing of ion-implanted
layers et cetera. In particular, to verify this microscopic
mechanism, a simulation of boron redistribution during
low-temperature annealing of ion-implanted layers has been carried
out under different conditions of transient enhanced diffusion
suppression. Due to the good agreement with the experimental data,
the values of the average migration length of nonequilibrium
impurity interstitials have been obtained. It has been shown that
for boron implanted into a silicon layer preamorphized by
germanium ions the average migration length of impurity
interstitials at the annealing temperature of 800 $^{\circ}$C can
be reduced from 11 nm to approximately 6 nm due to additional
implantation of nitrogen. The further shortening of the average
migration length is observed if the processing temperature is
reduced to 750 $^{\circ}$C. It is also found that for implantation
of $\mathrm{BF_{2}}$ ions into silicon crystal, the value of the
average migration length of boron interstitials is equal to 7.2 nm
for thermal treatment at a temperature of 800 $^{\circ}$C.
\end{abstract}

\pacs{61.72.Ji, 61.72.Tt, 61.72.Ss, 85.40.Ry}

\keywords{diffusion; annealing; doping effects; boron; silicon}


\maketitle

\section{Introduction}
It is well known that boron is a basic impurity of p-conductivity
used in the technology of the production of silicon integrated
microcircuits \cite{Salvador-10}. Unfortunately, boron atoms have
a small mass and large mobility in silicon crystals. Due to the
small masses of boron ions, the formation of an amorphous layer in
the ion-implanted substrates is not observed even at large
fluencies. Only a great number of radiation defects are created.
The absence of an amorphous layer and the presence of radiation
defects result in the significant transient enhanced diffusion
(TED) of ion-implanted boron during the subsequent annealing (see,
for example,
\cite{Hofker-75,Michel-87,Fan-87,Michel-1987,Eaglesham-94,Stolk-97,
Lerch-99,Ohuchi-01,Mirabella-02,Jain-02,Shao-03,Lerch-05,Graoui-05,
Gable-05,Lallement-05,Bazizi-10,Woon-10,Philippe-11,Shimizu-11}).
All these phenomena substantially complicate the problem of the
formation of very shallow junctions with high electrophysical
parameters. For suppressing the TED of ion-implanted boron, a
method of boron implantation in a silicon layer preamorphized by
heavier germanium ions is widely used
\cite{Ohuchi-01,Lerch-05,Graoui-05,Gable-05,Mirabella-04,
Camillo-Castillo-2004,Cristiano-04,Hopstaken-04,Pawlak-04,
Pawlak-004,Pawlak-2004,Camillo-Castillo-04,Tomita-04,Lindsay-04,
Hamilton-07,Ferri-07,Yeong-08}. Due to the solid phase epitaxial
regrowth (SPER) of the amorphous layer, the region doped with
boron is characterized by a perfect crystal structure, containing
defects that are invisible by electron microscopy. However, the
transient enhanced diffusion is observed as before, although it
has another character and a smaller intensity. In the papers
\cite{Velichko-10,Velichko-11} a qualitative difference of the
form of boron profiles produced by annealing at temperatures of
800 $^{\circ}$C and below was pointed out in comparison with the
annealing at 900 $^{\circ}$C and higher temperatures. Really, at
low annealing temperatures an extended ``tail'' is observed in the
low-concentration region of the impurity profile and the shape of
this ``tail'' is a straight line if the axis of concentration is
logarithmic. This feature of the boron distribution is observed
even for ``tail'' extension compared to the characteristic size of
the implanted region or smaller than it. At the same time, after
annealing at a temperature of 900 $^{\circ}$C or higher the shape
of the boron profile for concentrations below approximately
10$^{8}$ $\mu$m$^{-3}$ becomes convex upwards, i.e., similar to
the Gaussian distribution. This behavior of the profile shape
gives clear evidence of the change in the boron diffusion
mechanism.

It is worth noting that the ``tails'' that represent a straight
line, if the axis of concentration is logarithmic, are often
observed directly after ion implantation of boron, phosphorus,
gallium, and other impurity at room temperature (see, for example,
\qquad \cite{Hofker-75,Tomita-04} in the case of implantation of
boron ions, \qquad \cite{Moline-71,Blood-74} for phosphorus
implantation, \qquad \cite{Crowder-71,Dearnaley-75} in the case of
implantation of gallium ions). The ion implantation in all the
cases investigated was carried out in the direction deflecting
from the crystal axis. According to \cite{Ryssel-78}, to
completely remove the phenomenon of channeling and to eliminate
the ``tails'' related to the scattering of ions into channels, the
ion implantation should be carried out in amorphous silicon.
Nevertheless, later experiments show that ``tails'' are observed
for boron implantation in the layers preamorphized by germanium
ions \cite{Cristiano-04,Hamilton-07}.

For the low-temperature treatments of ion-implanted layers either
an increase and broadening of existing ``tails'' in the bulk of a
semiconductor occur or the formation of new ``tails'' if they are
not observed after implantation. In the region of low impurity
concentration, a ``tail'' represents as before a straight line for
low-temperature annealing (i.e. for a small thermal budget).
Really, an increase in the ``tail'' extension occurs during
subsequent thermal treatments of preamorphized silicon layers that
were implanted with boron ions
\cite{Cristiano-04,Pawlak-2004,Hamilton-07,Ferri-07}. In the
investigations carried out by
\cite{Gable-05,Camillo-Castillo-2004,Camillo-Castillo-04,Yeong-08}
clearly identified ``tails'' after boron implantation were not
observed. However, such ``tails'' were formed in the course of the
subsequent annealing. As follows from the experimental data of
\cite{Gamo-73}, a ``tail'' is also formed in the case of thermal
treatment at 900 $^{\circ}$C of silicon layers implanted with
indium. This ``tail'' represents a straight line if the
concentration axis is logarithmic. The experimental data also show
that ``tails'' characterized by a straight line are often observed
for the ``hot'' ion implantation of indium \cite{Gamo-73}, gallium
\cite{Gamo-73}, antimony \cite{Gamo-70}, and other impurities.

It was assumed originally that the formation of ``tails'' in the
low-concentration region of ion-implanted impurity profiles,
especially in the case of ``hot'' ion implantation, results from
the fast diffusion of implanted impurity atoms
\cite{Crowder-71,Gamo-73,Gamo-70,Davies-68}. It was supposed that
impurity interstitials are this fast diffusing species
\cite{Gamo-73,Crowder-71,Davies-68}. For example, it was shown
experimentally in \cite{Davies-68} that during annealing of
ion-implanted layers a significant fraction of indium and
tellurium atoms leave their substitutional positions and become
interstitials. The interstitial position is also a characteristic
feature for atoms of gallium that, as well as boron, indium, and
tellurium is the element of III groups. However, in the latest
papers \cite{Blood-74,Dearnaley-75} there are very serious
arguments that the formation of ``tails'' in ion-implanted layers
is related to the scattering of ions that reserved a part of the
kinetic energy into channels. For example, in the paper
\cite{Blood-74} phosphorus ions were implanted into thin silicon
layers of different thicknesses. These layers were located on a
substrate, which collected ions channeling through the layer. It
was assumed that diffusing atoms do not have sufficient energy to
leave the silicon crystal, whereas the channeling ions have.
Experiments showed that the substrate really collects ions passed
through the silicon layer. The doses of the passed ions were
obtained as a function of the layer thicknesses. These doses
correspond to the doses of phosphorus atoms in the region of the
remainder of the ``tail'' for the investigated depth if ions were
implanted in the continuous silicon. The results obtained were
generalized in \cite{Dearnaley-75} for the cases of indium and
gallium implantation. According to \cite{Dearnaley-75} the
radiation-enhanced diffusion of gallium is impossible at room
temperature. On the other hand, a characteristic ``tail'' is
observed experimentally due to the scattering of ions into
channels. Taking into account the possible annealing of damages
and recovery of the crystal structure in the region of the end of
the ion range, one can explain the results of \cite{Gamo-73} for
``hot'' high fluence implantation of indium and gallium ions
without attracting the concept of an anomalous diffusion. Besides,
at low temperatures a ``tail'' can be formed as a result of the
channeling of a part of impurity atoms at the initial stage of
implantation, when the amorphous phase was not formed as yet.
Nevertheless, the mechanism of the ``tail'' formation is not clear
until now, especially taking into account the latest experiments
\cite{Cristiano-04,Hamilton-07} related to boron implantation in
the layer preamorphized by implantation of germanium ions. This
allows us to formulate the following purpose of the research.

\section{\textbf{Main goal of the research}}
{\bf Not rejecting the possibility for a part of the ions of
scattering into channels, we are to show that the long-range
migration of nonequilibrium impurity interstitials is the main
factor in the formation of ``tails'' in the region of low impurity
concentration for random ion implantation into silicon crystals
and implantation into preamorphized silicon layers.}

\section{\textbf{Analysis of the mechanisms of the ``tail'' formation during ion implantation }}

Let us consider four characteristic cases of the formation of
``tails'' in the region of impurity concentration decreasing in
the bulk of the semiconductor:

(i) ``Tail'' formation during the subsequent annealing of
ion-implanted layers
\cite{Gamo-73,Gable-05,Camillo-Castillo-2004,Camillo-Castillo-04,Yeong-08};

(ii) Formation of ``tails'' during ion implantation into amorphous
silicon \cite{Cristiano-04,Hamilton-07};

(iii) The phenomenon of ``tail'' formation during ``hot'' ion
implantation \cite{Gamo-73,Gamo-70};

(iv) Formation of ``tails'' during ion implantation in the
direction deflecting from the axis of the crystal at room
temperature of the substrate
\cite{Hofker-75,Tomita-04,Moline-71,Blood-74,Crowder-71,Dearnaley-75}.

\begin{figure}[ht]
\centering {
\begin{minipage}[ht]{8.6 cm}
{\includegraphics[ scale=0.8]{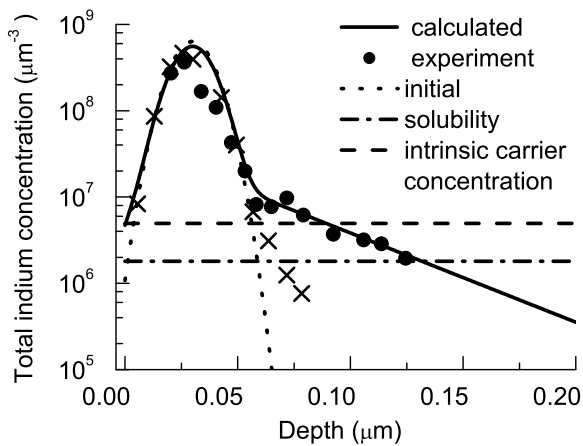}}
\end{minipage}}
\caption{Calculated indium concentration profile (solid line)
after thermal treatment of implanted silicon substrate at 900
$^{\circ}$C for 20 minutes. The experimental data ($\times$ --
impurity distribution after implantation; $\bullet$ -- after
annealing) are taken from \cite{Gamo-73} and the indium solubility
limit is taken from \cite{Solmi-02}}. \label{fig:Indium900}
\end{figure}

It is evident that in the first two cases, there is no phenomenon
of channeling and it is possible to explain the formation of
``tails'' only by the anomalous impurity diffusion. It is worth
noting that impurity diffusion can occur in an amorphous phase
too, especially, if the high-current implanter is used for ion
implantation and there is a possibility of substrate heating, as
it was the case in the experiments of Cristiano et al.
\cite{Cristiano-04} As an example, Fig.~\ref{fig:Indium900}
presents the calculation of ion-implanted indium redistribution.
The indium concentration profile was calculated within the
framework of the model for diffusion of impurity interstitials
described below. For comparison, the experimental data of
\cite{Gamo-73} are used. In the work of Gamo et al. \cite{Gamo-73}
the distributions of impurity atoms were obtained by measuring the
$\gamma$-ray intensities in a combination with the layer removal
technique. Indium was implanted with an energy of 45 keV to a dose
of $\sim$1$\times$10$^{15}$ ion/cm$^{2}$ and 8$^{\circ}$ off the
$<111>$  axis in order to reduce channeling effects. The
temperature of annealing was 900 $^{\circ}$C, and the thermal
treatment duration was 20 minutes. The following values of
parameters for the model of interstitial diffusion of
ion-implanted impurity were used. The parameters of the
as-implanted indium distribution are: $R_{p}$ = 0.03 $\mu$m (30
nm); $\Delta R_{p}$ = 0.0084 $\mu$m (8.4 nm); the parameters of
the indium interstitial diffusion are: the average migration
length of indium interstitials $l_{AI}$ = 0.042 $\mu$m (42 nm);
the time-average value of the generation rate of nonequilibrium
impurity interstitials in the maximum of distribution $g^{AI}_{m}$
= 7.9$\times$10$^{4}$ $\mu$m$^{-3}$s$^{-1}$. Here $R_{p}$ and
$\Delta R_{p}$ are the average projective range of impurity ions
and straggling of the projective range, respectively. These
parameters provide the best fitting of the calculated indium
concentration profile to the experimental one.

As can be seen from Fig.~\ref{fig:Indium900}, the results of
calculation agree well with the experimental data. It is a very
important argument in favour of the fast diffusion of
nonequilibrium indium interstitials.

\begin{figure}[ht]
\centering {
\begin{minipage}[ht]{8.6 cm}
{\includegraphics[ scale=0.74]{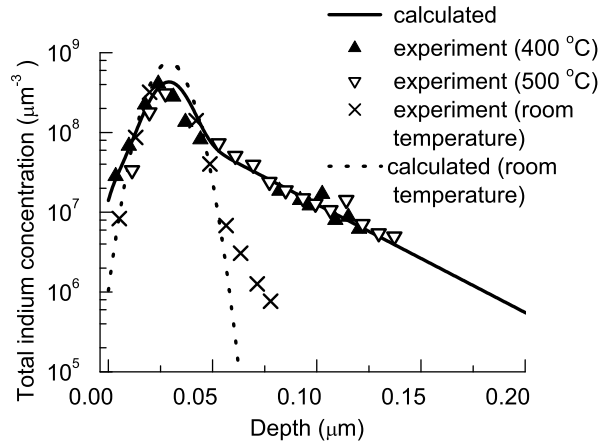}}
\end{minipage}}
\caption{Calculated indium concentration profile (solid line)
after ``hot'' ion implantation. The experimental data are taken
from \cite{Gamo-73}}. \label{fig:IndiumHot}
\end{figure}

Taking into account the good agreement with the experimental data,
we use the model of the interstitial diffusion of ion-implanted
impurity for simulating indium redistribution during ``hot'' ion
implantation that was investigated in \cite{Gamo-73}. The results
of simulation are presented in Fig.~\ref{fig:IndiumHot}. The
energy and the dose of indium ions are the same as in the previous
experiment, namely, 45 keV and $\sim$1$\times$10$^{15}$
ions/cm$^{2}$. The following value of the average migration length
of indium interstitials was used for the best fit to the
experimental data: $l_{AI}$ = 0.032 $\mu$m (32 nm). This value is
greater than the average migration length $l_{AI}$ = 0.027 $\mu$m
used for similar calculations in \cite{Velichko-88}. Perhaps, the
increase of $l_{AI}$ is due to the effect of evaporation of indium
interstitials from the surface of the semiconductor
\cite{Kizilyalli-96} that we took into consideration in contrast
to \cite{Velichko-88}. Taking into account the evaporation of
indium interstitials made it possible to achieve the best
agreement with the experimental profile in the vicinity of the
surface.

As can be seen from Fig.~\ref{fig:IndiumHot}, the results of
calculation agree well with experimental data. The value of the
average migration length for ``hot'' implantation of indium ions
$l_{AI}$  = 0.032 $\mu$m is close to the migration length $l_{AI}$
= 0.042 $\mu$m for annealing of indium implanted layer at a
temperature of 900 $^{\circ}$C. It means that in the experiment
with ``hot'' ion implantation a ``tail'' is also formed due to the
fast migration of impurity interstitials.

The results obtained suggested to revise the inferences drawn in
\cite{Blood-74,Dearnaley-75} about ``tail'' formation due to the
scattering of ions into channels. Really, these inferences were
based on the experiments \cite{Blood-74} with phosphorus ion
implantation in thin silicon layers of different thicknesses. The
main assumption of the above-mentioned paper is that diffusing
phosphorus atom cannot pass through the boundary of the silicon
crystal, while the ions which scatter into channels can cross the
boundary due to the rest of the kinetic energy. However, within
the framework of the interstitial diffusion model the ``tail''
formation is related to the diffusion of impurity interstitials.
Therefore, there are no restrictions for impurity interstitials to
evaporate from the surface of a crystal and then to collect on the
substrate, on which the implanted layer is placed. This is
confirmed by the experiments concerning hydrogen diffusion through
thin silicon layers \cite{Wieringen-56}, because the extended
``tails'' on the hydrogen concentration profiles are also formed
by interstitial diffusion \cite{Velichko-07}.

It is clear from the analysis carried out above that the vast
majority of phenomena of ``tail'' formation in the
low-concentration region of ion-implanted impurity profile is due
to the fast impurity diffusion. The scattering of ions into
channels can play only an auxiliary role in the experiments under
investigation. It is usually supposed that impurity interstitials
are the fast diffusing species
\cite{Gamo-73,Crowder-71,Davies-68,Velichko-2010}. Let us note
that the mechanism of the long-range migration of nonequilibrium
impurity interstitials was already used for explanation of gold
diffusion in ion-implanted layers \cite{Schulz-74}. This mechanism
was also used for describing ``tail'' formation during thermal
treatments of the layers implanted with boron
\cite{Velichko-83,Velichko-2010} and for explanation of boron
diffusion in ``buried'' layers under the conditions of generation
of nonequilibrium silicon interstitials
\cite{Cowern-90,Cowern-91}. Really, in \cite{Cowern-90} an
analytical solution describing the redistribution of impurity
atoms due to diffusion of boron interstitials during thermal
treatment was obtained. As the initial condition for the
distribution of boron atoms the $\delta$-function was chosen. It
was shown that for single participation of a boron atom in the act
of interstitial migration a ``tail'' represents a straight line if
the axis of concentration is logarithmic. It is necessary to note
that for the initial boron distribution in the form of
$\delta$-function, the requirement of the occurrence of one event
of boron interstitial migration is equivalent to the condition of
the long-range migration of boron interstitials. On the other
hand, under the condition of multiple migrations of the same
impurity atom the boron profile has the form of Gaussian
distribution \cite{Cowern-90}. Therefore, it was proposed in
\cite{Cowern-91} to determine the microscopic mechanism of
impurity transport by means of a change in the form of the
impurity profile with increase in annealing duration. It is worth
noting that the calculations which were carried out in
\cite{Velichko-10,Velichko-11,Velichko-07,Velichko-2010,Velichko-2011}
show that the ``tail'' representing a straight line is formed by
the long-range migration of impurity interstitials not only in the
case of initial distribution in the form of $\delta$-function, but
also for a generation rate of nonequilibrium interstitials
described by the Gaussian distribution or any other distribution.
Besides, the average migration length of impurity interstitials
can be comparable with the characteristic size of the initial
doped region, or even less than this size.

\section{\textbf{Model of interstitial diffusion}}
In the paper \cite{Velichko-08}, an equation for migration of
nonequilibrium impurity interstitials was obtained which takes
into account the different charge states of migrating species and
also the drift of impurity interstitials in the built-in electric
field and in the field of the elastic stresses. This equation has
a form similar to the form of the equation describing the
diffusion of nonequilibrium ``impurity atom -- intrinsic point
defect'' pairs when the influence of the built-in electric field
and the field of the elastic stresses are also taken into account.
As the mathematical formulations for the migration of
nonequilibrium impurity interstitials and for the diffusion due to
the nonequilibrium pairs are identical, it is impossible to make
an exact conclusion about the kind of the mobile species from the
form of impurity profile after thermal treatment. Let us consider
for definiteness that boron transport during low-temperature
annealing is carried out by migration of nonequilibrium impurity
interstitials. We shall neglect the diffusion of the ``impurity
atom -- intrinsic point defect'' pairs in view of the small
thermal budget. This assumption is based on the well-known opinion
that impurity interstitials are the most mobile species. Taking
into account that the extended ``tail'' is formed in the
low-concentration region, where $C\leq n_{i}$, we can suppose that
the electric field does not influence the migration of impurity
interstitials. Here $C$ is the concentration of the
substitutionally dissolved boron atoms; $n_{i}$ is the intrinsic
carrier concentration for the treatment temperature. It is clear
that there is also no influence of the electric field in the case
of migration of neutral interstitials even if $C>n_{i}$. We shall
also neglect the influence of elastic stresses and suppose that
substitutionally dissolved boron atoms and boron atoms
incorporated into clusters and extended defects are immobile.
Then, the system of equations describing redistribution of
ion-implanted boron during low-temperature annealing has the
following form:

{\bf (i). The conservation law for immobile impurity atoms:}

\begin{equation}\label{Conservation law}
\displaystyle \frac{\partial \, C^{T}(x,t)}{ \partial \, t} =
\displaystyle \frac{C^{AI}(x,t)}{\tau^{AI}} + G^{AS}(x,t)\, ,
\end{equation}

{\bf (ii). The equation of diffusion for nonequilibrium impurity
interstitials:}

\begin{equation}\label{nonequilibrium impurity interstitials}
d^{AI}\displaystyle \frac{\partial^{2} \, C^{AI}}{ \partial \,
x^{2}} - \displaystyle \frac{C^{AI}}{\tau^{AI}} + G^{AI}(x,t) = 0
\, ,
\end{equation}

or

\begin{equation}\label{Normalized equation}
- \left[ \displaystyle \frac{\partial^{2} \, C^{AI}}{ \partial \,
x^{2}} - \displaystyle \frac{C^{AI}}{l_{AI}^{2}}\right] =
\frac{\tilde{g}^{AI}(x,t)}{l_{AI}^{2}} \, ,
\end{equation}

where

\begin{equation}\label{Average migration lenght}
l_{AI}=\sqrt{d^{AI} \tau^{AI}} \, , \qquad \tilde{g}^{AI}(x,t)=
G^{AI}(x,t) \, \tau^{AI} \, .
\end{equation}

Here $C^{T}$ is the total concentration of substitutionally
dissolved impurity atoms and impurity atoms incorporated into
clusters or trapped by extended defects (immobile impurity atoms):
$C^{AI}$ is the total concentration of nonequilibrium impurity
interstitials in different charge states; $d^{AI}$ and $\tau^{AI}$
are the diffusivity and the average lifetime of nonequilibrium
impurity interstitials, respectively; $G^{AI}$ is the generation
rate of impurity interstitials. We use the stationary diffusion
equation for impurity interstitials in view of their large
migration length $l_{AI}$ ($l_{AI}\gg l_{fall}$, where $l_{fall}$
is the characteristic length of the decrease in the impurity
concentration in the high-concentration region of impurity
profile), and due to a small lifetime of these nonequilibrium
interstitials $\tau^{AI}$ ($\tau^{AI}\ll \tau_{p}$, where
$\tau_{p}$ is the duration of annealing). Let us note that the
system of equations (\ref{Conservation law}), (\ref{nonequilibrium
impurity interstitials}) is similar to the systems of equations
used in
\cite{Schulz-74,Velichko-83,Velichko-2010,Cowern-90,Cowern-91} for
the description of interstitial diffusion.

To describe the spatial distribution of impurity atoms after
implantation and spatial distribution of the generation rate of
boron interstitials, the Gaussian distribution is chosen:

\begin{equation}\label{Generation}
C_{0}(x)=C(x,0)= C_{m}\exp \left[ -\frac{(x-R_{p})^{2}}{2\triangle
R_{p}^{ \,2}}\right] \, ,
\end{equation}

\begin{equation}\label{Interstitial_Generation}
G^{AI}(x,t)= g^{AI}_{m}\exp \left[
-\frac{(x-R_{p})^{2}}{2\triangle R_{p}^{ \,2}}\right] \, ,
\end{equation}

where

\begin{equation}\label{Maximum_Concentration}
C_{m}=\displaystyle \frac{Q}{\sqrt{2\pi}\Delta R_{p}}\times
10^{-8} \quad [\mu m^{-3}] \, .
\end{equation}

Here $C_{m}$ is the maximal concentration of impurity atoms after
implantation; $g^{AI}_{m}$ is the maximal value of generation rate
of impurity interstitials per unit volume; $Q$ is the dose of ion
implantation [ion/cm$^{2}$]; $R_{p}$ and $\triangle R_{p}$ are the
average projective range of impurity ions and straggling of the
projective range, respectively.

It is clear from expressions (\ref{Generation}) and
(\ref{Interstitial_Generation}) that the model presupposes that
the generation rate of boron interstitials is proportional to the
total concentration of impurity atoms. Really, the analysis
carried out in \cite{Velichko-2010} shows that the boron
interstitials can be generated during thermal treatment not only
as a result of annealing of radiation defects, but also due to
dissolution or rearrangement of the clusters that incorporated
impurity atoms and also as a result of elastic stresses arising
due to the small atomic radius of boron. This assumption is
confirmed by the experimental data of \cite{Gouye-10}, where the
long-range migration of boron interstitials with the formation of
the characteristic ``tail'' was observed during doping of the
amorphous silicon layer deposited with the low thermal budget on a
silicon substrate and then subjected to solid phase epitaxy. Thus,
it follows from \cite{Gouye-10} that the long-range migration of
boron interstitials is observed in the case of no radiation
defects. As both the cluster concentration and intensity of
elastic stresses are proportional to the total concentration of
impurity atoms, the use of expression (6) for describing the
generation rate distribution of boron interstitials is quite
reasonable.

\section{\textbf{Results of simulation of ion-implanted boron
redistribution}}

In the case of the constant coefficients   and   the system of
equations (\ref{Conservation law}) and (\ref{Normalized equation})
allows obtaining an analytical solution. Such solutions for
different boundary conditions on a finite interval $[0, x_{F}]$
within the assumption of continuous generation of impurity
interstitials in the doped layer were obtained in
\cite{Velichko-07,Velichko-2011}. We used these solutions for
modeling the redistribution of ion-implanted boron. Let us note
that the diffusivity $d^{AI}$ has a constant value, if the
migration of boron interstitials is not influenced by the built-in
electric field. The average lifetime of these interstitials   has
a constant value in the case of absorption by the unsaturated
traps distributed homogeneously in the bulk of a semiconductor.
For example, such case of absorption is realized if replacement of
the silicon atom by the boron interstitial from the lattice site
to the interstitial one occurs.

\begin{figure}[ht]
\centering {
\begin{minipage}[ht]{8.6 cm}
{\includegraphics[ scale=0.8]{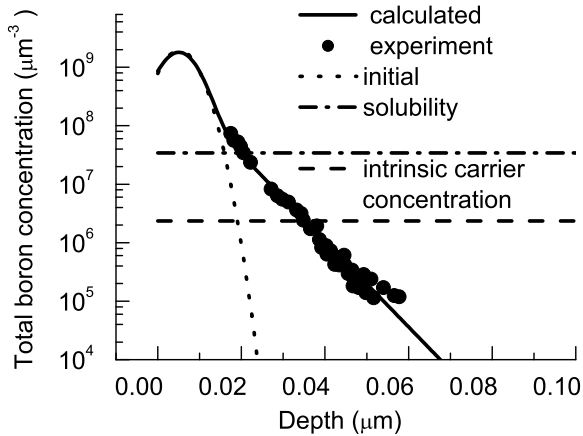}}
\end{minipage}}
\caption{Calculated boron concentration profile in silicon
preamorphized by Ge ions after annealing for 60 s at a temperature
of 800 $^{\circ}$C. The dotted curve represents calculated boron
distribution after ion implantation at an energy of 1 keV. An
additional implantation of nitrogen ions at an energy of 6 keV for
stronger suppression of the transient enhanced diffusion was
carried out. The experimental data (filled circles) are taken from
the paper \cite{Yeong-08}}. \label{fig:BoronNitrogen800}
\end{figure}

Let us consider in the beginning the results of simulation of the
experimental data of \cite{Yeong-08}, because in this
investigation various methods of the transient enhanced diffusion
suppression were used. So, in \cite{Yeong-08} for suppression of
the transient enhanced diffusion Czochralski grown (100) $n$-type
silicon wafers were subjected to preamorphization by performing a
germanium (Ge) ion implantation at 15 keV (Ge-PAI) and
subsequently with 1 keV boron ions to a dose of 1.5$\times
10^{15}$ cm$^{-2}$. Nitrogen (N) co-implantation to the same dose
of 1.5$\times 10^{15}$ cm$^{-2}$ was performed on some wafers. The
thermal annealing was carried out in a rapid thermal processing
system under N$_{2}$ ambient duration at a temperature of 800
$^{\circ}$C for 60 seconds. The dopant profiles were analyzed
ex-situ by secondary ion mass spectrometry (SIMS). A primary
Cross-sectional TEM (XTEM) was also performed to analyze the
extent of amorphization and the Ge-PAI induced EOR defects. The
simulation of boron redistribution carried out for these
conditions in \cite{Velichko-11}, has shown good agreement with
the experimental data for the average migration length of boron
interstitials $l_{AI}$ = 11 nm. Also it was supposed that
approximately 8.6\% of the implanted boron atoms occupied
interstitial positions. Migration of these nonequilibrium
interstitial atoms results in the formation of an extended
``tail'' on the boron concentration profile. This ``tail'' is
located in the interval from 0.02 $\mu$m up to approximately 0.1
$\mu$m from the surface of a semiconductor. For comparison
Fig.~\ref{fig:BoronNitrogen800} presents the results of modeling
for the same process of ion-implanted boron redistribution in the
case of additional nitrogen implantation at 6 keV after
preamorphization due to germanium ions \cite{Yeong-08}. The energy
of nitrogen implantation equal to 6 keV was chosen so that the
maximum of N concentration was located between the boron doped
region and a/c interface.

As can be seen from Fig.~\ref{fig:BoronNitrogen800}, there is good
agreement between the calculated profile and the experimental
data. The following values of the model parameters were used to
provide the best fit of the calculated boron concentration profile
to the experimental one: {\bf Parameters prescribing the initial
distribution of implanted boron:} $R_{p}$ = 0.0044 $\mu$m (4.4
nm); $\Delta R_{p}$ = 0.0036 $\mu$m (3.6 nm). {\bf Parameters
specifying the process of interstitial diffusion:} the duration of
annealing $\tau_{p}$ = 60 s; the annealing temperature T = 800
$^{\circ}$C; the average lifetime of nonequilibrium boron
interstitials $\tau^{AI}$ = 0.011 s; the maximum value of the
generation rate of nonequilibrium impurity interstitials
$g^{AI}_{m}$ = $6.7\times 10^{6} \quad \mu$m$^{-3}$s$^{-1}$; the
average migration length of boron interstitials $l_{AI}$ = 5.9 nm;
the concentration of boron interstitials on the right boundary
$C^{AI}_{F}$ = 0; the position of the right boundary $x_{F}$ = 0.5
$\mu$m. It was supposed that approximately 21 \% of the implanted
boron atoms occupied the interstitial positions temporally.

As can be seen from the calculations performed, the additional
implantation of nitrogen ions results in a significant (almost 2
times) reducing of the average migration length of boron
interstitials. Therefore, the shrinkage of the ``tail''
approximately by 0.03 $\mu$m is observed that is very attractive
from the technological point of view. It is worth noting that
there is a significant (2.8 times) increase in the generation rate
of boron interstitials. However, due to the reduced migration
length of these interstitials, their main fraction is trapped
within the implanted layer that does not result in an increase of
the depth of $p-n$ junction. We assume that this trapping is due
to the interaction with nitrogen atoms or with complexes of
nitrogen and germanium atoms.

\begin{figure}[ht]
\centering {
\begin{minipage}[ht]{8.6 cm}
{\includegraphics[ scale=0.8]{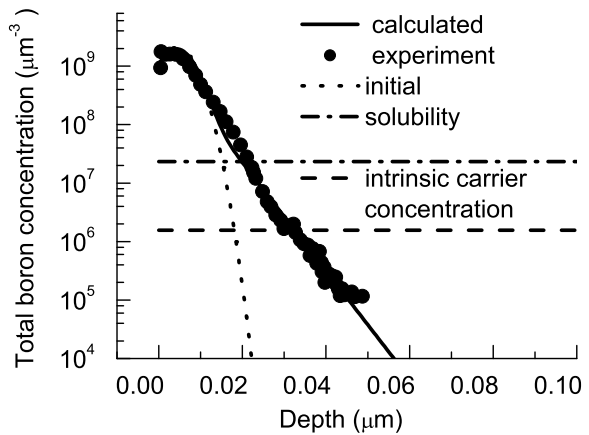}}
\end{minipage}}
\caption{Calculated boron concentration profile in silicon
preamorphized by Ge ions after annealing for 60 s at a temperature
of 750 $^{\circ}$C. The dotted curve represents calculated boron
distribution after ion implantation at an energy of 1 keV. An
additional implantation of nitrogen ions at an energy of 6 keV for
stronger suppression of the transient enhanced diffusion was
carried out. The experimental data (filled circles) are taken from
the paper \cite{Yeong-08}}. \label{fig:BoronNitrogen750}
\end{figure}

A similar modeling for the case of transient enhanced diffusion
suppression by using the lower annealing temperature, namely 750
$^{\circ}$C, is presented in Fig.~\ref{fig:BoronNitrogen750}. {\bf
The following values of the parameters specifying the process of
interstitial diffusion were used:} the maximum value of the
generation rate of nonequilibrium impurity interstitials
$g^{AI}_{m}$ = $7.4\times 10^{6} \quad \mu$m$^{-3}$s$^{-1}$; the
average migration length of boron interstitials $l_{AI}$ = 4.7 nm.
It follows from the values obtained that the decrease of the
annealing temperature results in the reduction of the average
migration length of boron interstitials and in the highly abrupt
impurity profile.

\begin{figure}[ht]
\centering {
\begin{minipage}[ht]{8.6 cm}
{\includegraphics[ scale=0.8]{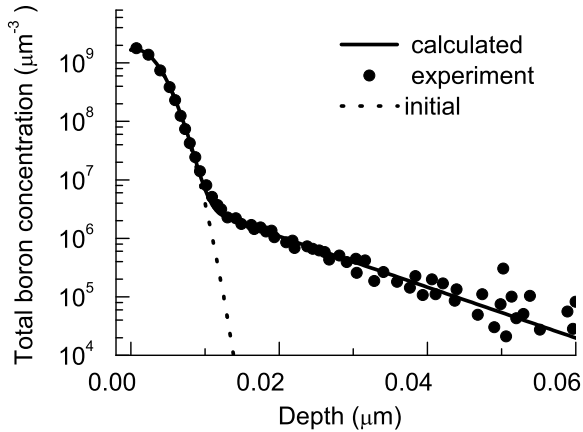}}
\end{minipage}}
\caption{Calculated concentration profile of boron implanted into
silicon preamorphized by Ge ions. The long-range migration of
boron atoms during implantation is taken into account. The dotted
curve represents boron distribution after ion implantation at an
energy of 0.5 keV calculated with no interstitial diffusion. The
experimental data (filled circles) are taken from
\cite{Cristiano-04}}. \label{fig:BoronImplant}
\end{figure}

Now, let us present the results of modeling the interstitial
diffusion during boron implantation using the data obtained in
\cite{Cristiano-04}. In the paper \cite{Cristiano-04}, $n$-type
10-20 $\Omega$cm, prime (100) Si wafers were used. The wafers were
first preamorphized with Ge ions (30 keV, $1\times 10^{15} \quad
$cm$^{-2}$) to a depth of about 50 nm and then implanted with 0.5
keV boron ions to a dose of $1\times 10^{15} \quad $cm$^{-2}$. The
implants were performed on a high-current implanter at tilt and
twist angles of 0$^{\circ}$ with electrostatic deceleration in
front of the target. It is worth noting that due to the
high-current implanter the wafer can be heated during ion
implantation. The dopant atom distributions of all samples were
analyzed using a high-resolution secondary ion mass spectrometry
(SIMS). The boron concentration profile measured after
implantation in preamorphized silicon is shown in
Fig.~\ref{fig:BoronImplant}.

As can be seen from Fig.~\ref{fig:BoronImplant}, the calculated
curve agrees well with the measured boron concentration profile
after implantation characterized by the extended ``tail''. Thus,
the experimental data \cite{Cristiano-04} can be explained on the
basis of the long-range  migration of nonequilibrium boron atoms.
Really, there is no channeling of ions due to implantation in the
amorphous phase, whereas fast anomalous diffusion is quite
possible in amorphous silicon. The following values of the model
parameters were used to provide the best fit of the calculated
boron concentration profile to the experimental one: {\bf
Parameters prescribing the Gaussian distribution of implanted
boron:} $R_{p}$ = 0.0006 $\mu$m (0.6 nm); $\Delta R_{p}$ = 0.0027
$\mu$m (2.7 nm). {\bf Parameters specifying the process of fast
diffusion:} the maximum value of the generation rate of
nonequilibrium impurity interstitials $g^{AI}_{m}$ = 3.2$\times
10^{5} \quad \mu$m$^{-3}$s$^{-1}$; the average migration length of
boron interstitials $l_{AI}$ = 0.01 $\mu$m (10 nm). For the best
fit it was supposed that approximately 0.63 \% of the implanted
boron atoms participated in the long-range migration. However,
there is also another way to explain the ``tail'' formation
\cite{Cristiano-11}. Really, the boron ions were implanted with
electrostatic deceleration in front of the target. During the
deceleration stage, some ions can become neutral due to reciprocal
collisions. In such a case they will not be decelerated and keep
their original energy. Due to the higher energy, these ions
penetrate deeper in the bulk of the semiconductor. Therefore, this
problem requires a further investigation.

\begin{figure}[ht]
\centering {
\begin{minipage}[ht]{8.6 cm}
{\includegraphics[ scale=0.8]{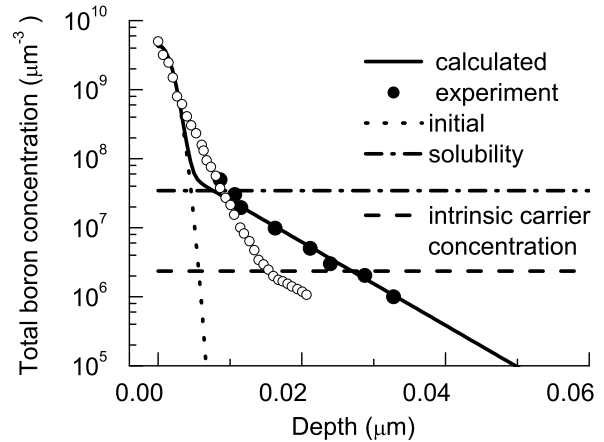}}
\end{minipage}}
\caption{Calculated boron concentration profile after annealing of
a silicon layer implanted with BF$_{2}$ ions for TED suppression.
The dotted curve represents the calculated boron distribution
after implantation of BF$_{2}$ ions. The annealing was carried out
for 30 minutes at a temperature of 800 $^{\circ}$C. The
experimental data (open circles after implantation and filled
circles after annealing) are taken from the paper by Suzuki
\cite{Suzuki-01}}. \label{fig:BF2-800}
\end{figure}

Finally, in Fig.~\ref{fig:BF2-800} the results of modeling of
ion-implanted boron redistribution are shown, when the
implantation of BF$_{2}$ ions was used for amorphization of the
surface layer of silicon crystal and transient enhanced diffusion
suppression \cite{Suzuki-01}. In \cite{Suzuki-01}, (100) oriented,
10 $\Omega$cm, $n$-type silicon wafers were subjected to BF$_{2}$
implantation at 2.2 keV to a dose of 1.0$\times 10^{15} \quad
$cm$^{-2}$. The furnace annealing was carried out at a temperature
of 800 $^{\circ}$C for 30 minutes. The resulting boron
concentration profiles were measured by SIMS. The following values
of the model parameters were used to provide the best fit of the
calculated boron concentration profile to the experimental one:
{\bf Parameters prescribing the initial distribution of implanted
boron:}  $R_{p}$ = 0.0002 $\mu$m (0.2 nm); $\Delta R_{p}$ = 0.0014
$\mu$m (1.4 nm). {\bf Parameters specifying the process of
interstitial diffusion:} the maximum value of the time-average
generation rate of nonequilibrium impurity interstitials
$g^{AI}_{m}$ = 2.0$\times 10^{5} \quad \mu$m$^{-3}$s$^{-1}$; the
average migration length of boron interstitials $l_{AI}$ = 0.0072
$\mu$m (7.2 nm). It was supposed that approximately 4.1 \% of the
implanted boron atoms occupied the interstitial positions
temporally.

As is seen from Fig.~\ref{fig:BF2-800}, the calculated boron
profile after furnace annealing agrees well with the experimental
one in the ``tail'' region of impurity distribution that look like
a straight line. It means that the ``tail'' region is formed due
to the long-range migration of nonequilibrium boron interstitials.
It is necessary to note that modeling of the experimental data
obtained in \cite{Suzuki-01} was carried out earlier in the paper
\cite{Marcon-08}. The model of Uematsu \cite{Uematsu-97} was used
for simulation of boron diffusion in \cite{Marcon-08}. It is
supposed in this model that there is a local thermodynamic
equilibrium between substitutionally dissolved boron atoms,
silicon self-interstitials, and boron interstitials. It means from
the microscopic point of view that the same boron atom becomes
frequently interstitial, makes a number of chaotic jumps, and then
becomes substitutionally dissolved again due to the "kick-out" of
silicon atom from the lattice site. According to \cite{Cowern-91},
the impurity profile for this diffusion mechanism has a form of
the Gaussian distribution. Really, the boron concentration profile
calculated in \cite{Marcon-08} represents the Gaussian
distribution and disagrees with the experimental data. On the
other hand, as follows from the experimental data
\cite{Suzuki-01}, the boron concentration profile after annealing
for 1.5 h begins to get a convex form similar to the Gaussian
distribution. Remaining within the framework of the model of the
long-range migration, such a change in the form of impurity
distribution can be explained by taking into account the
significant increase in the boron concentration in the ``tail''
region. Really, if the impurity concentration is greater than the
intrinsic carrier concentration $n_{i}$, the migration of charged
boron interstitials is influenced by the built-in electric field.
On the other hand, the increase of annealing duration can create
appropriate conditions for intensive dissolution of the clusters
and extended defects incorporating silicon self-interstitials. In
turn the generation of nonequilibrium self-interstitials results
in boron diffusion due to the generation, migration, and
dissolution of the ``boron atom -- silicon self-interstitial''
pairs. It is usually supposed (see, for example,
\cite{Velichko-84,Morehead-86,Mathiot-91,Bracht-07}) that the
local thermodynamic equilibrium prevails between substitutionally
dissolved boron atoms, silicon self-interstitials, and the pairs.
Therefore, the boron concentration profile formed due to the pair
diffusion should look similar to the Gaussian distribution that
agrees with the experimental data for large thermal budget.
However, the final solution of this problem requires further
careful investigations.

\section{\textbf{Conclusions}}
We have shown that the vast majority of the phenomena of ``tail''
formation in the low-concentration region of ion-implanted
impurity profile, including ion implantation in the preamorphized
silicon, ``hot'' ion implantation, impurity redistribution during
the subsequent low-temperature annealing of ion-implanted layers,
is related to the fast impurity diffusion. The channeling of ions
scattered into channels is negligible for all the cases under
consideration. With this purpose, simulation of indium
redistribution during the subsequent annealing of ion-implanted
layers and during ``hot'' indium implantation has been carried
out. Many cases of the ``tail'' formation under different
conditions of transient enhanced diffusion suppression during
low-temperature annealing of boron implanted layers have been also
investigated, including boron implantation into silicon
preamorphized by germanium ions. The impurity concentration
profiles calculated within the framework of the model of the
long-range migration of nonequilibrium impurity interstitials
agree well with experimental data. On this basis the values of the
average migration length of nonequilibrium impurity interstitials
have been obtained. So, if for boron implantation in silicon
crystals preamorphized by germanium ions, the average migration
length of boron interstitials is equal to 11 nm at an annealing
temperature of 800 $^{\circ}$C, the additional nitrogen
implantation allows one to reduce this value approximately to 6
nm. A further reduction of the average migration length of boron
interstitials can be achieved by decreasing the annealing
temperature to 750 $^{\circ}$C. Finally, for BF$_{2}$ implantation
in silicon crystals the average migration length of boron
interstitials is equal to 7.2 nm at an annealing temperature of
800 $^{\circ}$C.


\begin{thebibliography}{99}

\bibitem{Salvador-10}
D. De Salvador, E. Napolitani, G. Bisognin, M. Pesce, A. Carnera,
E. Bruno, G. Impellizzeri, and S. Mirabella, Boron diffusion in
extrinsically doped crystalline silicon. Phys. Rev. B. {\bf
Vol.81}, Art.No. 045209 (2010).

\bibitem{Hofker-75}
W.K. Hofker Implantation of boron in silicon. Philips Res. Repts.
Suppl. No.8. pp.1-121 (1975).

\bibitem{Michel-87}
A. E. Michel, W. Rausch, P. A. Ronsheim, R. E. Kastl, Rapid
annealing and the anomalous diffusion of ion implanted boron into
silicon. Appl. Phys. Lett. {\bf Vol.50}, No.7. pp.416-418 (1987).

\bibitem{Fan-87}
D. Fan, J. Huang, R. J. Jaccodine, P. Kahora, F. Stevie, Enhanced
tail diffusion of ion implanted boron in silicon. Appl. Phys.
Lett. {\bf Vol.50}, No.24. pp.1745-1747 (1987).

\bibitem{Michel-1987}
A. E. Michel, W. Rausch, P. A. Ronsheim, Implantation damage and
the anomalous diffusion of ion-implanted boron. Appl. Phys. Lett.
{\bf Vol.51}, No.7. pp.487-489. (1987).

\bibitem{Eaglesham-94}
D. J. Eaglesham, P. A. Stolk, H.-J. Gossmann, J. M. Poate,
Implantation and transient B diffusion in Si: The source of the
interstitials. Appl. Phys. Lett. {\bf Vol.65}, No.18. pp.2305-2307
(1994).

\bibitem{Stolk-97}
P. A. Stolk, H.-J. Gossmann, D. J. Eaglesham, D. C. Jacobson, C.
S. Rafferty, G. H. Gilmer, M. Jaraiz, J. M. Poate, H. S. Luftman,
T. E. Haynes, Physical mechanisms of transient enhanced dopant
diffusion in ion-implanted silicon. J. Appl. Phys. {\bf Vol.81},
No.9. pp.6031-6050 (1997).

\bibitem{Lerch-99}
W. Lerch, M. Gl\"{u}ck, N. A. Stolwijk, H. Walk, M. Sch\"{a}fer,
S. D. Marcus, D. F. Downey, J. W. Chow, Boron ultrashallow
junction formation in silicon by low-energy implantation and rapid
thermal annealing in inert and oxidizing ambient. J. Electrochem.
Soc. {\bf Vol.146}, No.7. pp.2670-2678 (1999).

\bibitem{Ohuchi-01}
K. Ohuchi, K. Adachi, A. Murakoshi, A. Hokazono, T. Kanemura, N.
Aoki, M. Nishigohri, K. Suguro, Y. Toyoshima, Ultrashallow
junction formation for sub-100 nm complementary
metal-oxide-semiconductor field-effect transistor by controlling
transient enhanced diffusion. Jpn. J. Appl. Phys. {\bf Vol.40},
Pt.1, No.4B. pp.2701-2705 (2001).

\bibitem{Mirabella-02}
S. Mirabella, A. Coati, S. Scalese, D. De Salvador, S. Pulvirenti,
G. Bisognin, E. Napolitani, A. Terrasi, M. Berti, A. Carnera, A.
V. Drigo, F. Priolo, Suppression of boron transient enhanced
diffusion by c trapping. Solid State Phenom. {\bf Vols.82-84}.
pp.195-200 (2002).

\bibitem{Jain-02}
S. C. Jain, W. Schoenmaker, R. Lindsay, P. A. Stolk, S. Decoutere,
M. Willander, H. E. Maes, Transient enhanced diffusion of boron in
Si. J. Appl. Phys. {\bf Vol.91}, No.11. pp.8919-8940 (2002).

\bibitem{Shao-03}
L. Shao, J. Liu, Q. Y. Chen, Wei-Kan Chu, Boron diffusion in
silicon: the anomalies and control by point defect engineering.
Mat. Sci. Eng. R. {\bf Vol.42}. pp.65-114 (2003).

\bibitem{Lerch-05}
W. Lerch, S. Paul, J. Niess, S. McCoy, T. Selinger, J. Gelpey, F.
Cristiano, F. Severac, M. Gavelle, S. Boninelli, P. Pichler, D.
Bolze, Advanced activation of ultra-shallow junctions using
flash-assisted RTP. Mat. Sci. Eng. B. {\bf Vols.124-125}. pp.24-31
(2005).

\bibitem{Graoui-05}
H. Graoui, M. A. Foad, A comparative study on ultra-shallow
junction formation using co-implantation with fluorine or carbon
in pre-amorphized silicon. Mat. Sci. Eng. B. {\bf Vols.124-125}.
pp.188-191 (2005).

\bibitem{Gable-05}
K. A. Gable, L. S. Robertson, A. Jain, K. S. Jones, The effect of
preamorphization energy on ultrashallow junction formation
following ultrahigh-temperature annealing of ion-implanted
silicon. J. Appl. Phys. {\bf Vol.97}, Art.No. 044501 (2005).

\bibitem{Lallement-05}
F. Lallement, D. Lenoble, Investigation on boron transient
enhanced diffusion induced by the advanced P$^{+}$/N ultra-shallow
junction fabrication processes. Nuclear Instrum. Methods Phys.
Res. Sect. B. {\bf Vol.237}. pp.113-120 (2005).

\bibitem{Bazizi-10}
E. M. Bazizi, P. F. Fazzini, A. Pakfar, C. Tavernier, B. Vandelle,
H. Kheyrandish, S. Paul, W. Lerch, F. Cristiano, Modeling of the
effect of the buried Si-SiO$_{2}$ interface on transient enhanced
boron diffusion in silicon on insulator. J. Appl. Phys. {\bf
Vol.107}, Art.No. 074503 (2010).

\bibitem{Woon-10}
W. Y. Woon, C. L. Chen, Impact of ion implantation boundary
dimensionality on boron transient diffusion in submicron scale
patterns. Appl. Phys. Lett. {\bf Vol.97}, Art.No. 121907 (2010).

\bibitem{Philippe-11}
T. Philippe, S. Duguay, D. Mathiot, D. Blavette, Atomic scale
evidence of the suppression of boron clustering in implanted
silicon by carbon coimplantation. J. Appl. Phys. {\bf Vol.109},
Art.No. 023501 (2011).

\bibitem{Shimizu-11}
Y. Shimizu, H. Takamizawa, K. Inoue, T. Toyama, Y. Nagai, N.
Okada, M. Kato, H. Uchida, F. Yano, T. Tsunomura, A. Nishida, T.
Mogami, Impact of carbon coimplantation on boron behavior in
silicon: Carbon-boron coclustering and suppression of boron
diffusion. Appl. Phys. Lett. {\bf Vol.98}, Art.No. 232101 (2011).


\bibitem{Mirabella-04}
S. Mirabella, D. De Salvador, E. Napolitani, F. Giannazzo, G.
Impellizzeri, G. Bisognin, A. Terrasi, V. Raineri, M. Berti, A.
Carnera, A.V. Drigo, F. Priolo, Self-interstitial diffusion and
clustering with impurities in crystalline silicon. Nuclear
Instrum. Methods Phys. Res. Sect. B. {\bf Vol.216}. pp.80-89
(2004).

\bibitem{Camillo-Castillo-2004}
R. A. Camillo-Castillo, M. E. Law, K. S. Jones, Impact of dopant
profiles on the end of range defects for low energy germanium
preamorphized silicon. Mat. Sci. Eng. B {\bf Vols.114-115}.
pp.312-317 (2004).

\bibitem{Cristiano-04}
F. Cristiano, N. Cherkashin, P. Calvo, Y. Lamrani, X. Hebras, A.
Claverie, W. Lerch, S. Paul, Thermal stability of boron electrical
activation in preamorphized ultra-shallow junctions. Mater. Sci.
Eng. B. {\bf Vols.114-155}. pp.174-179 (2004).

\bibitem{Hopstaken-04}
M. J. P. Hopstaken, Y. Tamminga, M. A. Verheijen, R. Duffy, V. C.
Venezia, A. Heringa, Effects of crystalline regrowth on dopant
profiles in preamorphized silicon. Appl. Surf. Sci. {\bf
Vol.231-232}. pp.688-692 (2004).

\bibitem{Pawlak-04}
B. J. Pawlak, R. Surdeanu, B. Colombeau, A. J. Smith, N. E. B.
Cowern, R. Lindsay, W. Vandervorst, B. Brijs, O. Richard, F.
Cristiano, Evidence on the mechanism of boron deactivation in
Ge-preamorphized ultrashallow junctions. Appl. Phys. Lett. {\bf
Vol.84}, No.12. pp.2055-2057 (2004).

\bibitem{Pawlak-004}
B. J. Pawlak, W. Vandervorst, A. J. Smith, N. E. B. Cowern, B.
Colombeau, X. Pages, Enhanced boron activation in silicon by high
ramp-up rate solid phase epitaxial regrowth. Appl. Phys. Lett.
{\bf Vol.86}, Art.No. 101913 (2004).

\bibitem{Pawlak-2004}
B. J. Pawlak, R. Lindsay, R. Surdeanu, B. Dieu, L. Geenen, I.
Hoflijk, O. Richard, R. Duffy, T. Clarysse, B. Brijs, W.
Vandervorst, C. J. J. Dachs, Chemical and electrical dopants
profile evolution during solid phase epitaxial regrowth. J. Vac.
Sci. Technol. B. {\bf Vol.22}, No.1. pp.297-301 (2004).

\bibitem{Camillo-Castillo-04}
R. A. Camillo-Castillo, M. E. Law, K. S. Jones, L. M. Rubin,
Influence of low temperature preanneals on dopant and defect
behavior for low energy Ge preamorphized silicon, J. Vac. Sci.
Technol. B. {\bf Vol.22}, No.1. pp.312-316 (2004).

\bibitem{Tomita-04}
M. Tomita C. Hongo, M. Suzuki, M. Takenaka, A. Murakoshi,
Ultra-shallow depth profiling with secondary ion mass
spectrometry. J. Vac. Sci. Technol. B. {\bf Vol.22}, No.1. pp.
317- 322 (2004).

\bibitem{Lindsay-04}
R. Lindsay, K. Henson, W. Vandervorst, K. Maex, B. J. Pawlak, R.
Duffy, R. Surdeanu, P. Stolk, J. A. Kittl, S. Giangrandi, X.
Pages, K. van der Jeugd, Leakage optimization of ultra-shallow
junctions formed by solid phase epitaxial regrowth, J. Vac. Sci.
Technol. B. {\bf Vol.22}, No.1.  pp. 306-311 (2004).

\bibitem{Hamilton-07}
J. J. Hamilton, K. J. Kirkby, N. E. B. Cowern, E. J. H. Collart,
M. Bersani, D. Giubertoni, S. Gennaro, A. Parisini, Boron
deactivation in preamorphized silicon on insulator: Efficiency of
the buried oxide as an interstitial sink, Appl. Phys. Lett. {\bf
Vol.91}, Art.No. 092122 (2007).

\bibitem{Ferri-07}
M. Ferri, S. Solmi, D. Giubertoni, M. Bersani, J. J. Hamilton, M.
Kah, K. Kirkby, E. J. H. Collart, N. E. B. Cowern, Uphill
diffusion of ultralow-energy boron implants in preamorphized
silicon and silicon-on-insulator. J. Appl. Phys. {\bf Vol.102},
Art.No. 103707 (2007).

\bibitem{Yeong-08}
S. H. Yeong, B. Colombeau, K. R. C. Mok, F. Benistant, C. J. Liu,
A. T. S. Wee, G. Dong, L. Chan, M. P. Srinivasan, The impact of
nitrogen co-implantation on boron ultra-shallow junction formation
and underlying physical understanding. Mat. Sci. Eng. B. {\bf
Vols.154-155}. pp.43-48 (2008).

\bibitem{Velichko-10}
O. I. Velichko, A. A. Hundorina, Modeling of the redistribution of
boron atoms during annealing of silicon layers produced by high
dose ion implantation. in: V. B. Odzhaev, V. V. Petrov, V. A.
Pilipenko et al. (Eds.), Proceedings of IV International Conf.
``Materials and Structures of Modern Electronics'', September,
23-24, 2010, Minsk, Belarus.  pp.112-115 (2010) (In Russian).

\bibitem{Velichko-11}
O. I. Velichko, A. A. Hundorina, Change in the microscopic
diffusion mechanisms of boron implanted into silicon with increase
in the annealing temperature.
arXiv:1105.4270v1[cond-mat.mtrl-sci].

\bibitem{Moline-71}
R. A. Moline, Ion-implanted phosphorous in silicon: profiles using
C-V analysis. J. Appl. Phys. {\bf Vol.42}. pp.3553-3558 (1971).

\bibitem{Blood-74}
P. Blood, G. Dearnaley, M. A. Wilkins, The origin of non-Gaussian
profiles in phosphorus-implanted silicon. J. Appl. Phys. {\bf
Vol.45}, No.12. pp.5123-5128 (1974).

\bibitem{Crowder-71}
Crowder B. L. The influence of the amorphous phase on ion
distributions and annealing behavior of group III and group IV
ions implanted into Si. J. Electrochem. Soc. {\bf Vol.118}, No.6.
pp.943-952 (1971).

\bibitem{Dearnaley-75}
G. Dearnaley, G. A. Gard, W. Temple, M. A. Wilkins, Depth
distribution of gallium ions implanted into silicon crystals.
Appl. Phys. Lett. {\bf Vol.27}, No.1. pp.17-18 (1975).

\bibitem{Ryssel-78}
H. Ryssel, I. Ruge. {\it Ion Implantation} (John Wiley and Sons
Inc; 99th edition, 1986) 478 pages.

\bibitem{Gamo-73}
K. Gamo, M. Iwaki, K. Masuda, S. Namba, S. Ishihara, I. Kimura, I.
V. Mitchell, G. Ilic, J. L. Whitton, J. A. Davies, Enhanced
diffusion and lattice location of indium and gallium implanted in
silicon. Jpn. J. Appl. Phys. {\bf Vol.12}, No.5. pp.735-741
(1973).

\bibitem{Gamo-70}
K. Gamo, K. Masuda, S. Namba, S. Ishihara, I. Kimura, Enhanced
diffusion of high-temperature ion-implanted antimony into silicon.
Appl. Phys. Lett. {\bf Vol.17}, No.9. pp.391-393 (1970).

\bibitem{Davies-68}
J. A. Davies, L. Eriksson, J. W. Mayer, Experimental evidence for
interstitial In and Tl in ion-implanted silicon. Appl. Phys. Lett.
{\bf Vol.12}, No.8. pp.255-256 (1970).

\bibitem{Solmi-02}
S. Solmi, A. Parisini, M. Bersani, D. Giubertoni, V. Soncini, G.
Carnevale, A. Benvenuti, A. Marmiroli, Investigation on indium
diffusion in silicon. Appl. Phys. {\bf Vol.92}, No.3. pp.1361-1366
(2002).

\bibitem{Velichko-88}
O. I. Velichko. {\it Atomic Diffusion Processes under
Nonequilibrium State of the Components in a Defect--Impurity
System of Silicon Crystals}. Ph.D. Dissertation. (Institute of
Electronics of the National Academy of Sciences of Belarus, Minsk,
1988) (In Russian).

\bibitem{Kizilyalli-96}
I. C. Kizilyalli, T. L. Rich, F. A. Stevie, C. S. Rafferty,
Diffusion parameters of indium for silicon process modeling. J.
Appl. Phys. {\bf Vol.80}, No.9. pp.4944-4947 (1996).

\bibitem{Wieringen-56}
A. Van Wieringen, N. Warmoltz, On the permeation of hydrogen and
helium in single crystal silicon and germanium at elevated
temperatures. Physica. {\bf Vol.22}. pp.849-865 (1956).

\bibitem{Velichko-07}
O. I. Velichko, N. A. Sobolevskaya, Analytical solutions for the
interstitial diffusion of impurity atoms. Nonlinear Phenom.
Complex Syst. {\bf Vol.10}, No.4. pp.376-384 (2007).

\bibitem{Velichko-2010}
O. I. Velichko, N. V. Kniazhava, Modeling of the long-range
interstitial migration of ion implanted boron. Comput. Mat. Sci.
{\bf Vol.48}. pp.409-412 (2010).

\bibitem{Schulz-74}
M. Schulz, A. Goetzberger, Controlled gold doping of silicon by
using ion implantation. Appl. Phys. {\bf Vol.3}. pp.275-280
(1974).

\bibitem{Velichko-83}
O. I. Velichko, Modelirovanie protsessa pereraspredeleniya
ionno-implantirovannoi primesi pri korotkih nizkotemperaturnyh
termoobrabotkah (Modeling of the process of ion implanted impurity
redistribution under short low temperature thermal treatments)
Abstracts of the 7 International Conf. on Ion Implantation in
Semiconductors and Other Materials (Vilnius, September 27-29,
1983). pp.198-199 (In Russian).

\bibitem{Cowern-90}
N. E. B. Cowern, K. T. F. Janssen, G. F. A. van de Walle, D. J.
Gravesteijn, Impurity diffusion via an intermediate species: The
B-Si system. Phys. Rev. Lett. {\bf Vol.65}, No.19. pp.2434-2437
(1990).

\bibitem{Cowern-91}
N. E. B. Cowern, G. F. A. van de Walle, D. J. Gravesteijn, C. J.
Vriezema, Experiments on atomic-scale mechanisms of diffusion.
Phys. Rev. Lett. {\bf Vol.67}, No.2. pp.212-215 (1991).

\bibitem{Velichko-2011}
O.I. Velichko, N.A. Sobolevskaya, Analytical solution of the
equations describing interstitial migration of impurity atoms,
Nonlinear Phenom. Complex Syst. {\bf Vol.14}, No.1. pp.70-79
(2011).

\bibitem{Velichko-08}
O. I. Velichko, Macroscopic description of the diffusion of
interstitial impurity atoms considering the influence of elastic
stress on the drift of interstitial species. Phil. Mag. {\bf
Vol.88}, No.10. pp.1477-1491 (2008).

\bibitem{Gouye-10}
A. Gouy\'{e}, I. Berbezier, L. Favre, M. Aouassa, G. Amiard, A.
Ronda, Y. Campidelli, A. Halimaoui, Insights into solid phase
epitaxy of ultrahighly doped silicon. J. Appl. Phys. {\bf
Vol.108}, Art.No. 013513 (2010).

\bibitem{Cristiano-11}
F. Cristiano, Private communication.

\bibitem{Suzuki-01}
K. Suzuki, High-concentration diffusion profiles of low-energy
ion-implanted B, As and BF$_{2}$ in bulk silicon. Solid-State
Electron. {\bf Vol.45}, No.10. pp.1747-1751 (2001).

\bibitem{Marcon-08}
J. Marcon, A. Merabet, Diffusion modelling of low-energy
ion-implanted BF$_{2}$ in crystalline silicon: Study of fluorine
vacancy effect on boron diffusion. Mat. Sci. Eng. B. {\bf
Vols.154-155}. pp.216-220 (2008).

\bibitem{Uematsu-97}
M. Uematsu, Simulation of boron, phosphorus, and arsenic diffusion
in silicon based on an integrated diffusion model, and the
anomalous phosphorus diffusion mechanism. J. Appl. Phys. {\bf
Vol.82}, No.5. pp. 2228-2246 (1997).

\bibitem{Velichko-84}
O. I. Velichko, A set of equations of radiation-enhanced diffusion
of ion-implanted impurities. in: I. I. Danilovich, A. G. Koval',
V. A. Labunov et al. (Eds.), Proceedings of VII International
Conf. ``Vzaimodeistvie Atomnyh Chastits s Tverdym Telom
(Interaction of Atomic Particles with Solid)'', Part 2, Minsk,
Belarus, 1984, pp.180-181 (in Russian).

\bibitem{Morehead-86}
F. F. Morehead, R. F. Lever, Enhanced ``tail'' diffusion of
phosphorus and boron in silicon: Self-interstitial phenomena.
Appl. Phys. Lett. {\bf Vol.48}, No.2. pp.151-153 (1986).

\bibitem{Mathiot-91}
D. Mathiot, S. Martin, Modeling of dopant diffusion in silicon: An
effective diffusivity approach including point-defect couplings.
J. Appl. Phys. {\bf Vol.70}, No.6. pp.3071-3080 (1991).

\bibitem{Bracht-07}
H. Bracht, Self- and foreign-atom diffusion in semiconductor
isotope heterostructures. I. Continuum theoretical calculations.
Phys. Rev. B. {\bf Vol.75}, Art.No. 035210 (2007).

\end{thebibliography}
\end{document}